\documentclass[aip,reprint]{revtex4-1}
\usepackage{graphicx}
\usepackage{amsmath}
\usepackage{graphicx}
\usepackage{dcolumn}
\usepackage{bm}
\usepackage{epsfig}
\usepackage{epstopdf}

\begin{document}


\title{Drop Traffic in Microfluidic Ladder Networks with Fore-Aft Structural Asymmetry}
\author{Jeevan Maddala, William S. Wang, Siva A. Vanapalli and Raghunathan Rengaswamy}
\affiliation{Department of Chemical Engineering, Texas Tech University, Lubbock, TX 79401-3121}



\date{\today}

\begin{abstract}
We investigate the dynamics of pairs of drops in microfluidic ladder networks with slanted bypasses, which break the fore-aft structural symmetry. Our analytical results indicate that unlike symmetric ladder networks, structural asymmetry introduced by a single slanted bypass can be used to modulate the relative drop spacing, enabling them to contract, synchronize, expand, or even flip at the ladder exit.  Our experiments confirm all these behaviors predicted by theory. Numerical analysis further shows that while ladder networks containing several identical bypasses are limited to nearly linear transformation of input delay between drops, combination of forward and backward slant bypasses can cause significant non-linear transformation enabling coding and decoding of input delays.

\end{abstract}

\maketitle

Understanding the spatiotemporal dynamics of confined immiscible plugs in interconnected fluidic paths is essential for applications ranging from lab-on-chip technologies \cite{review_angewad,song_ismagilov} to physiological flows \cite{vascular_flow_review} to  porous media flows \cite{prs_media}. The traffic of  drops or bubbles in even simple networks such as bifurcating channels can be astonishingly complex due to collective hydrodynamic resistive interactions  in the branches \cite{spack,drp_trfc_engl,fuerstman}. Although such intricate dynamics, in the case of lab-on-chip applications, make device design challenging, the collective behaviors can be harnessed to perform useful tasks such as droplet sorting \cite{cristobalsortin}  and storage \cite{swastika,rails_anchors_stroing}.

\indent Recently, the collective dynamics between pairs of drops have been harnessed in the so-called microfluidic ladder networks (MLNs) to control their relative drop spacing \cite{manu,*manu_conf}.  In MLNs, two droplet-carrying parallel channels are connected by narrow bypass channels through which the motion of drops is forbidden but the carrier fluid can leak. Current versions of ladders have fore-aft structural symmetry due to equally-spaced vertical bypasses. Such symmetric ladders are limited in functionality because the distance between pairs of drops have been shown to decrease at the exit only for constant inlet flow \cite{rail_road,manu,ajdari}. Since flexible manipulation of drop spacing in networks is crucial for passively regulating a variety of tasks including drop coalescence \cite{ladder_sync_drp_gen}, detection and storage, there is a need to design microfluidic ladders with multiple functionalities.

From a fundamental perspective, the dynamics of drops in MLNs is distinct compared to the widely-studied microfluidic loops \cite{spack,ajdari,fuerstman,cmptn_locl_flws}. In loops, drops at junctions choose a given branch. These discrete choices make such systems non-linear, enabling coding and decoding of input signals. Since drops do not typically make decisions at the bypass junctions in ladders, an open question is: is it possible to design microlfuidic ladders that yield significant non-linear transformation of input signal?

\indent In this Letter, we study the dynamics of spacing between drop pairs in MLNs with slanted bypasses. We find that because the slant breaks the fore-aft symmetry, it provides significantly more control over drop spacing than symmetric MLNs. We also discover that inclusion of slanted bypasses in ladders can non-linearly transform the initial delay between drops. These advanced capabilities arise because slanted bypasses flexibly manipulate \textit{(i)} the locations in the channels where drop velocity changes occur and \textit{(ii)} the time drops spend with bypasses between them.

\begin{figure}[ht]
 \includegraphics[height = 3.5 cm, width=8cm]{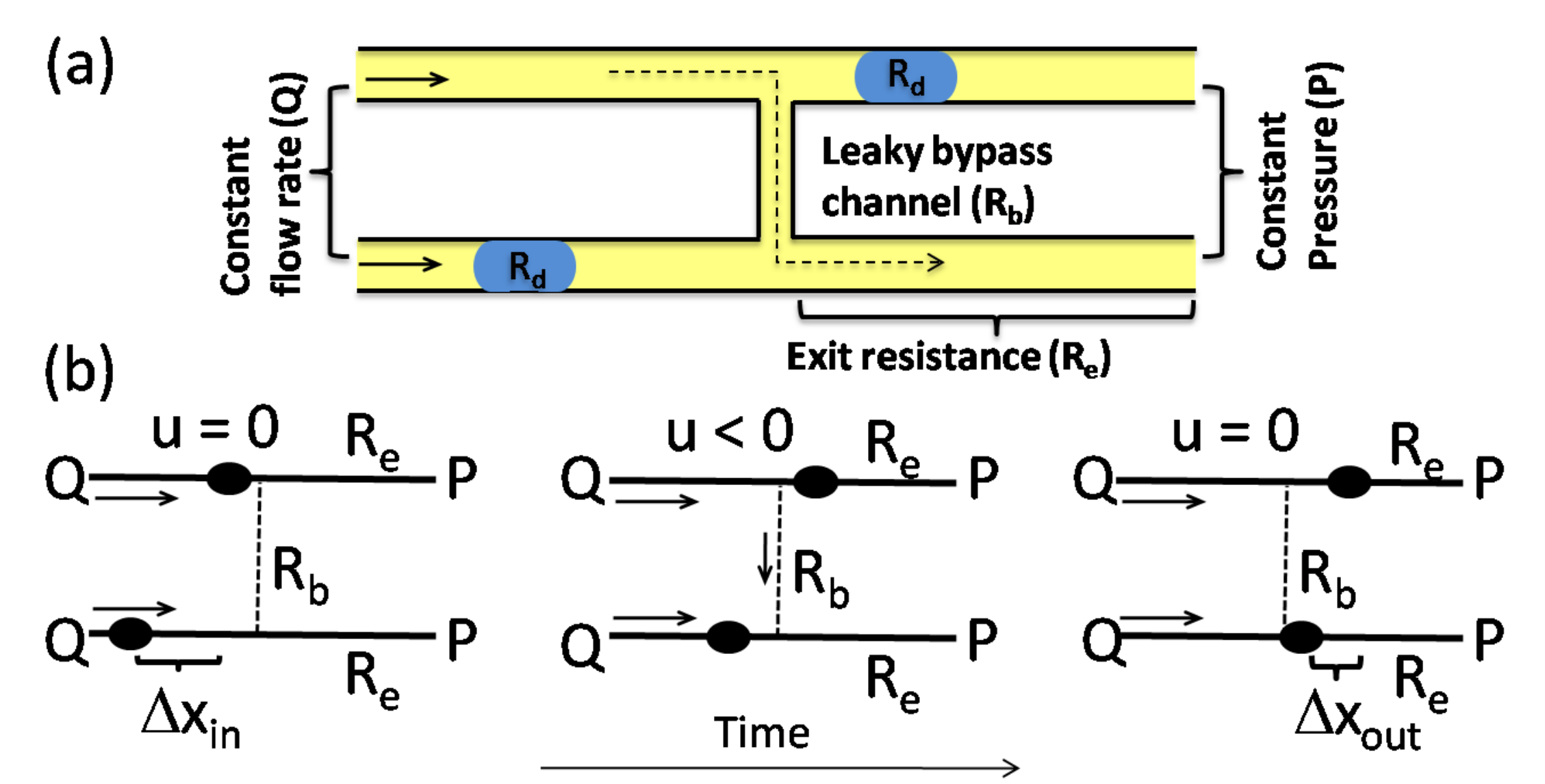}%
 \caption{(a) Ladder with a vertical bypass (b) Three distinct configurations are possible when a pair of drops traverse through a symmetric ladder network; Black objects represent drops. Full and dashed lines denote transport and bypass channels respectively. Arrows show flow direction.}
 \label{fig:confgr}
 \end{figure}

\indent As shown in Fig. \ref{fig:confgr}, the key  framework for quantifying drop spacing comes from understanding the variation in relative velocity ($u$) between drops in the top and bottom channel as they cross nodes in the network. Consider the simple case of a symmetric MLN with one vertical bypass. As shown in Fig. \ref{fig:confgr}(b) left, when two drops driven by a constant flow rate enter the ladder with an initial separation $(\Delta x_{in})$, they maintain the same separation, as $u=0$. When the leading drop crosses the node, a new configuration (see Fig. \ref{fig:confgr}(b) center) is reached, and the relative velocity changes as fluid leaks into the bypass, i.e. $u < 0$, causing a contraction in inter-drop distance. When both drops cross the bypass,  $u=0$ again as the pressure drop across the bypass is zero due to equal downstream branch resistances (see Fig. \ref{fig:confgr}(b) right). Thus, the outlet drop spacing, $(\Delta x_{out})$, is less than the initial separation and is given by $\Delta x_{out}$ = $\Delta x_{in}$+u$\Delta T$, where $\Delta T$ is the duration drops remain in the particular configuration of Fig. \ref{fig:confgr}(b) center.

In networks with many bypasses,  more  drop configurations and relative velocity changes are possible than that of Fig. \ref{fig:confgr}(b) as drops traverse through multiple nodes. In general, we find that
 \begin{equation}
 {\Delta}x_{out} = {\Delta}x_{in} + \displaystyle\sum_{j=1}^p u_j\Delta T_j
  \label{eqn:delxt}
  \end{equation}
  where $p$ is the number of distinct configurations of droplets occurring in the network, $u_j$ and $\Delta T_j$ are  the associated relative velocities and time periods.

\indent The degree of separation achieved at the ladder exit depends on the contribution of the summation term in Eqn (\ref{eqn:delxt}) both with respect to magnitude and sign. The strength of this contribution is modulated by the specific architecture of the ladder network. To compute this contribution, we use resistive network modeling \cite{ajdari}, where we assume each drop is a point object with the same hydrodynamic resistance $(R_d)$. Since the drop velocity $(V)$ is linearly dependent on liquid flow rate $(Q)$ \cite{garstecki_linear_flow}, we have $\mathfrak{V = \beta Q/S}$ , where $S$ is the channel cross-sectional area and $\mathfrak{0 < \beta < 2}$. To preserve the relative separation when drops leave the ladder network, we choose the downstream channel resistances to be equal.

\begin{figure}[ht]
 \includegraphics[height = 6cm, width=0.45\textwidth]{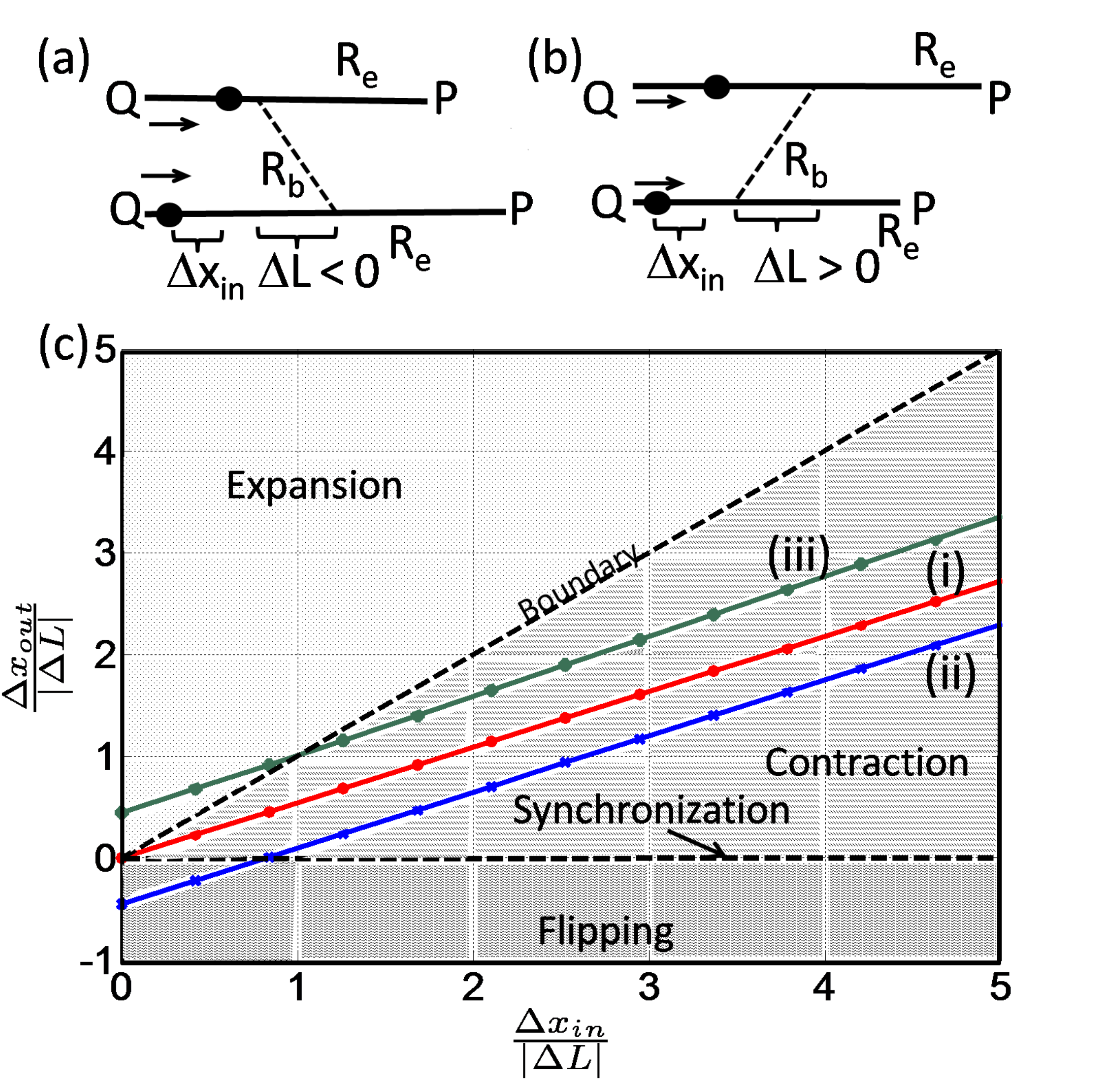}%
 \caption{Ladder networks with (a) backward slant (b) forward slant; (c) Dynamical regimes due to drop traffic in MLNs with a single (i) vertical bypass  (ii) backward slant and (iii) forward slant. M = 0.45.}
 \label{fig:basicunit}
 \end{figure}

\indent We begin by discussing the effect of a single slanted bypass in regulating the dynamics of drop spacing.  Fig. \ref{fig:basicunit} ($a$, $b$) shows a representative ladder network with a slanted bypass. In contrast to the vertical bypass, a new control parameter, $\Delta L$ is needed to describe two possible structural configurations---backward slant for $\Delta L < 0$ and forward slant for $\Delta L > 0$.

\indent To determine the drop spacing at the exit due to the slanted bypass, we identify the relative velocities that are non-zero and the corresponding durations as prescribed by Eqn (\ref{eqn:delxt}). Similar to the vertical bypass in Fig. \ref{fig:confgr}, non-zero $u$ occurs only after one of the droplets crosses a node. Solving the hydrodynamic circuit for this drop configuration analytically, we obtain  $\mathfrak{{u} = \beta Q/S \cdot (R_d/(R_{b}+2R_e+R_d))}$ and $\mathfrak{\Delta T = (|{\Delta}x_{in}-\Delta L|)S/({\beta}Q})$, where $R_b$ and $R_e$ are the bypass and exit channel resistance respectively. Thus, Eqn (\ref{eqn:delxt}) for the case of an MLN with a slanted bypass transforms to
\begin{eqnarray}
{\Delta}x_{out} = {\Delta}x_{in}-M ({\Delta}x_{in}-\Delta L)
\label{eqn:comprhnsve}
\end{eqnarray}
where $\mathfrak{M = {R_d}/{(R_b+2R_e+R_d)}}$ and 0 $<$ M $<$ 1. Note in Eqn (\ref{eqn:comprhnsve}), $\Delta x_{in} > 0$ corresponds to the top drop leading over the bottom drop in the ladder.

\indent Remarkably, Eqn (\ref{eqn:comprhnsve}) captures several dynamical regimes emerging from structural asymmetry due to a slanted bypass as  illustrated in Fig. \ref{fig:basicunit}(c). For the particular case of a vertical bypass ($\Delta L=0$), Eqn (\ref{eqn:comprhnsve}) reveals that drops can only undergo contraction (see Fig. \ref{fig:confgr}(a)). Perfect synchronization of drop pairs, i.e. $\Delta x_{out}$=0 is difficult to achieve with a single vertical bypass, as it requires $M$ to be unity.

\indent In contrast to the vertical bypass,  we find that the backward slant, where $\Delta L < 0$, yields flexible control over drop spacing as illustrated in Fig. \ref{fig:basicunit}(ii). For large input drop spacing, $\mathfrak{{\Delta}x_{in} > \frac{M\Delta L}{M-1}}$, the pairs at the exit undergo contraction for the same reason as in the vertical bypass. Perfect synchronization can also be realized with just a single backward slant when $\mathfrak{{\Delta}x_{in} = \frac{M\Delta L}{M-1}}$.  Moreover, when $\mathfrak{{\Delta}x_{in} < \frac{M\Delta L}{M-1}}$, a new  regime emerges that we refer to as $\textit flipping$. We observe that the leading droplet is initially ahead of the lagging droplet. However, when the leading drop crosses the bypass first, its velocity is reduced and the lagging drop has sufficient duration to catch up and overtake it. Thus, the flipping behavior yields $\Delta x_{out} < 0$ as shown in Fig. \ref{fig:basicunit}(ii).

\indent Similar to the backward slant, the forward slant (where $\Delta L > 0$) provides additional means of control. Interestingly, in contrast to the backward slant, the behavioral transitions depend only on the value of input drop spacing relative to $\Delta L$. When $\mathfrak{{\Delta}x_{in} < \Delta L}$, the lagging drop crosses the bypass first, resulting in expansion as highlighted in Fig. \ref{fig:basicunit}(iii). Alternatively, if $\mathfrak{{\Delta}x_{in} > \Delta L}$, the leading droplet crosses the bypass first, resulting in contraction. If $\mathfrak{{\Delta}x_{in} = \Delta L}$, then both drops cross the bypass simultaneously, and the input spacing is preserved. Thus forward slant allows drop pairs to expand, contract or remain unchanged.

\indent It would be misleading to visualize the slanted ladder network as being equivalent to a vertical bypass network with the addition of $\Delta L$ to the inlet spacing as we have already shown that vertical bypasses can only reduce the drop spacing for a fixed inlet flow. We assumed the exit channels to have equal lengths to simplify the analysis; however, the analysis presented in this letter extends trivially to unequal exit lengths also.

\begin{figure}[ht]
 \includegraphics[height = 6.81cm, width=8cm]{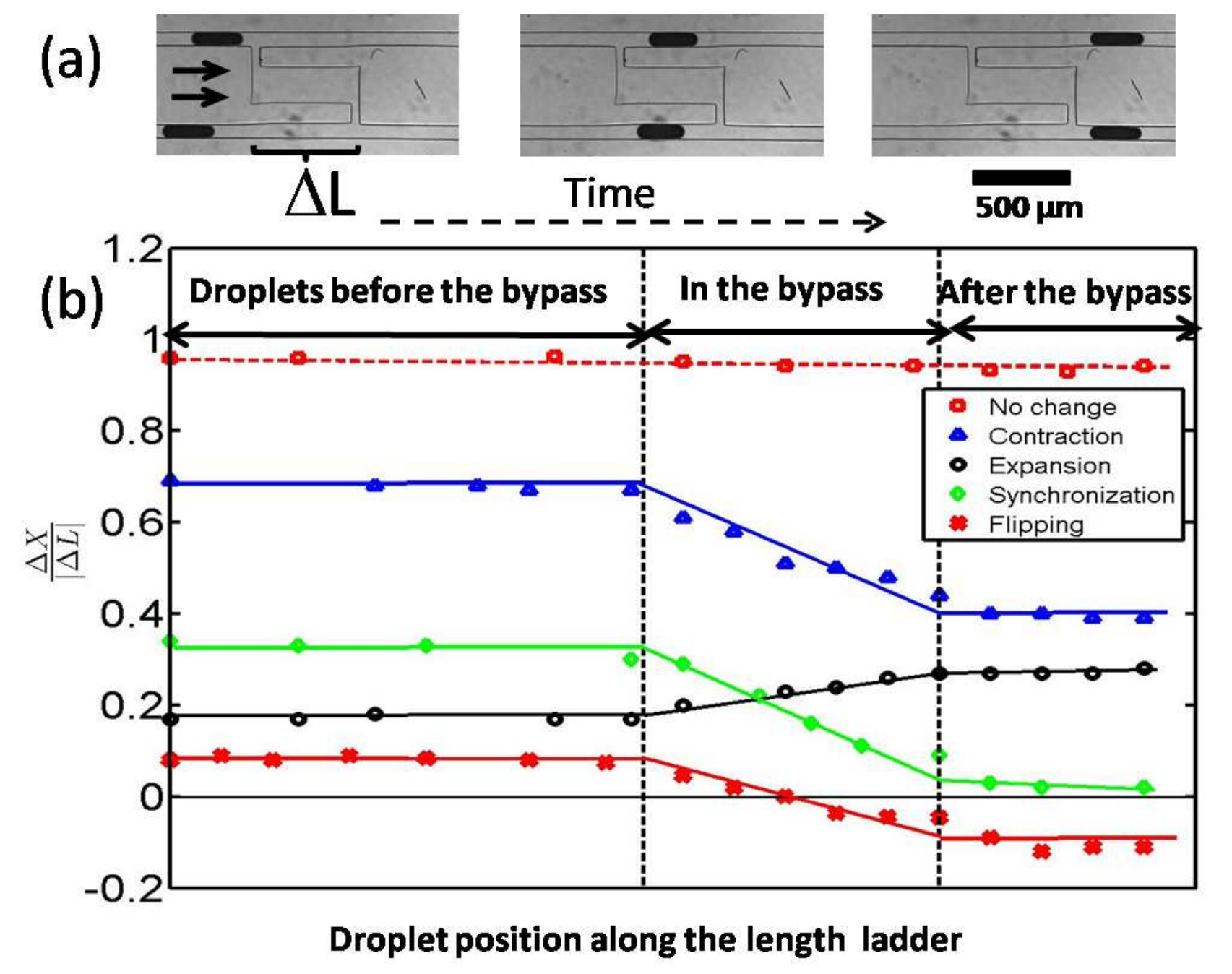}%
 \caption{Experimental confirmation of the dynamic behaviors in MLNs: (a) Snapshots showing synchronization of a droplet pair. Time interval between images is $0.08s$; (b) Drop spacing as a function of position in the ladder ($|\Delta L| = 500 \mu m$). Continuous phase was hexadecane, and dispersed phase was aqueous dye solution. The transport channels are $100 \mu m$ wide and tall.}
 \label{fig:experimental}
 \end{figure}

\indent To confirm the different behaviors predicted by our theory, we sought to construct an MLN with a single slant. However, the design space is large, requiring optimization of upstream and downstream transport channel resistance, slant resistance, slant slope, and hydrodynamic resistance of drops, which itself is a complex function of the flow conditions and fluid properties \cite{siva_drop_res,labrot}. Fortunately, insights from Eqn (\ref{eqn:comprhnsve}) reduce the search space. According to Fig. \ref{fig:experimental}, the backward slant is the best candidate to achieve maximum contraction, perfect synchronization and flipping. Moreover, Eqn (\ref{eqn:comprhnsve}) reveals that if $\Delta x_{in} < 0$ ($.i.e.,$ the top drop is lagging behind the bottom drop in the ladder), then the forward slant becomes a backward slant, allowing access to the expansion regime as well as the condition where the input separation does not change. Thus, we chose the backward slant to test our predictions.

\indent We incorporated two flow-focusing drop generators in polydimethyl(siloxane) devices to introduce drops at a constant flow rate into the ladder. To amplify the effects produced by the backward slant, we maximized the value of $M$ by minimizing $R_b$ and $R_e$. For example, as shown in Fig. \ref{fig:experimental}(a), the bypass has an enlarged mid-section to minimize $R_b$. We also ensured $R_b$ and $R_e$ to be $\sim$ O($R_d$). In Fig. \ref{fig:experimental}(b), we show experimental curves corresponding to each of the dynamical regimes predicted by our theory. We find the drop spacing to be relatively unchanged when the drop pair is before or after the bypass. However, as pairs of drops cross the bypass section, their spacing may expand, contract, flip, synchronize or remain unchanged depending on the initial separation. By comparing the data of Fig. \ref{fig:experimental} with the expression for $u$, we estimate  $R_d \approx 1.2kg/mm^{4}s$; $M \approx 0.2$. Taken together, our analytical results and experiments suggest that MLNs with slanted bypasses provide greater flexibility than ladders with vertical bypasses.

\begin{figure}[h]
 \includegraphics[height = 3.0cm, width=8cm]{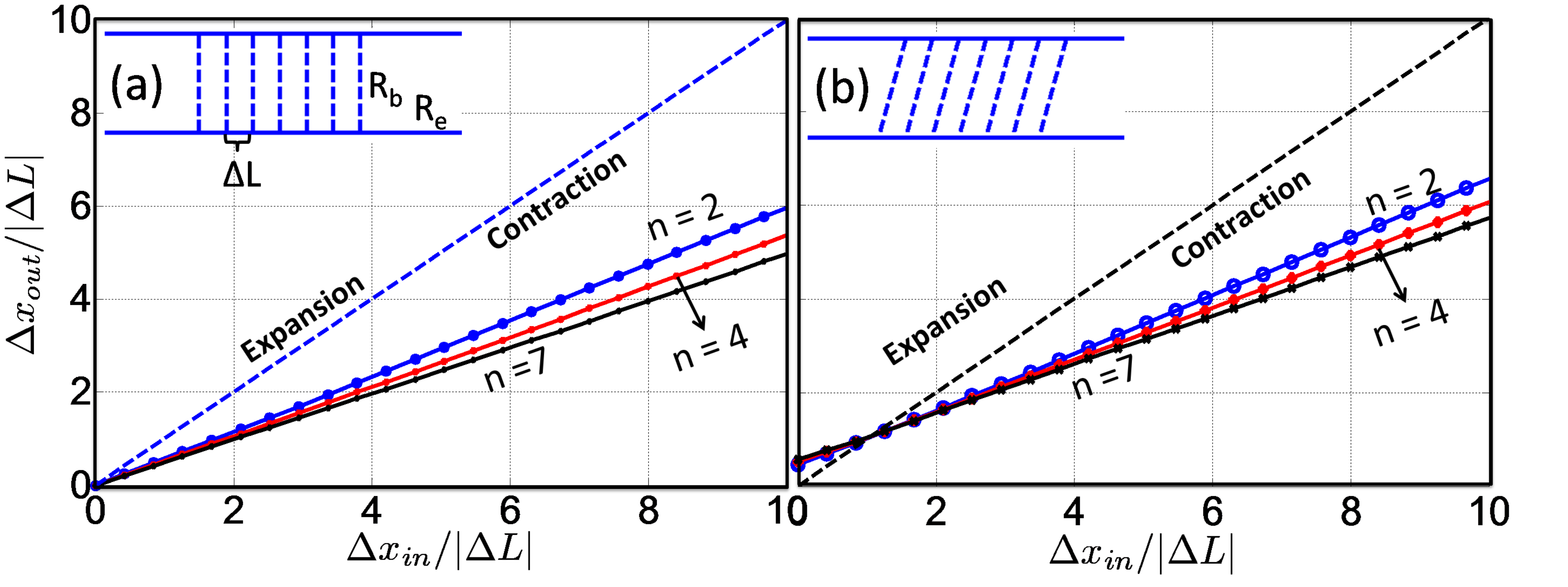}%
 \caption{Ladder networks with multiple identical bypasses: (a) Vertical bypasses (b) Forward slants. $\mathfrak{R_d/R_{b} = 3}$ ${\mathfrak{R_d/R_{\Delta L}} = 22}$,${\mathfrak{R_{d}/R_{e}} = 2}$,$\mathfrak{R_d = 1.5kg/mm^{4}s}$ and $\mathfrak{\beta = 1.4.}$}
\label{fig:slopeslant}
 \end{figure}

\indent To understand the flexibility due to additional bypasses, we investigated ladders containing several identical bypasses. In contrast to the single bypass case, in ladders with $n$ bypasses, the maximum number of configurations where $u$ is non-zero is $n(n+1)/2$. However, all these configurations need not be realized for a given $\Delta x_{in}$, which makes the theoretical analysis complex. We therefore used resistive network based simulations \cite{ajdari} to fully quantify the exit drop spacing for arbitrary input delay. As shown in Fig. \ref{fig:slopeslant}(a), we find that additional vertical bypasses simply amplify the contraction effect due to a single bypass, \textit{i.e.,} slope decreases with increasing number of bypasses.  A similar outcome also holds, as shown in Fig. \ref{fig:slopeslant}(b), for the particular case of an MLN with forward slants. We also find that the largest change in drop spacing occurs in the first few bypasses. Ladders with multi-bypasses could therefore be useful to maintain the same behavior as their single bypass counterparts, while dampening the effect of small fluctuations in input drop spacing.

\begin{figure}
 \includegraphics[height = 7.0cm, width=8cm]{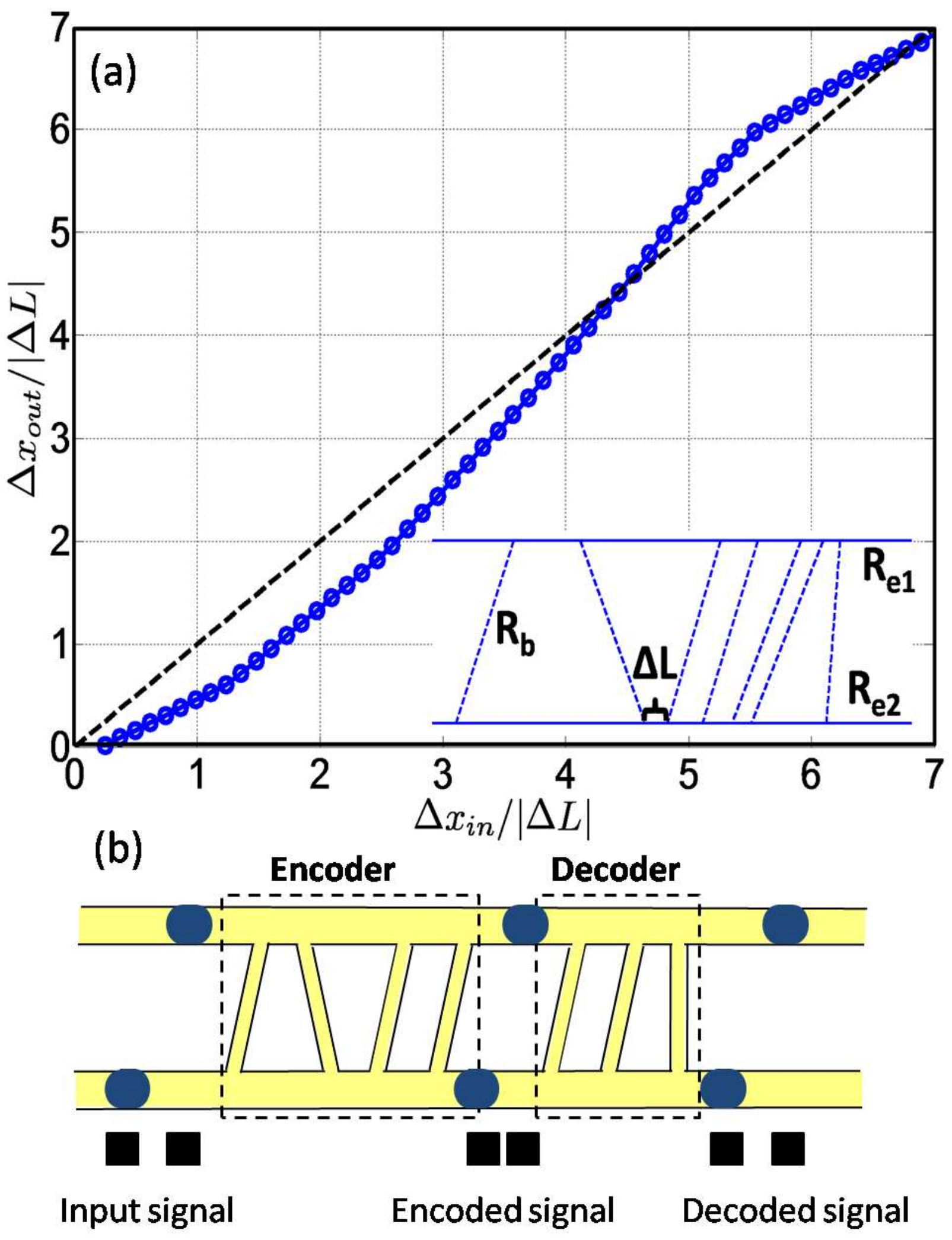}%
 \caption{ (a) Nonlinear output delay in a ladder network containing a mixed combination of slanted and vertical bypasses (structure of the network is shown in inset) (b) Encoding and decoding signal using ladder network  $\mathfrak{R_d/R_{b} = 3}$ ${\mathfrak{R_d/R_{\Delta L}} = 2.6}$, $(R_{e2})$ to $(R_{e1})$ is $1.01$,${\mathfrak{R_{e1}/R_{d}} = 20}$. $\mathfrak{R_d = 1.5kg/mm^{4}s}$ and $\mathfrak{\beta = 1.4.}$
\label{fig:nonlinear}}
 \end{figure}

\indent Our analysis of MLN designs with identical bypasses has shown that at small input delay, drop spacing may either contract or expand, while it always contracts at large input delay (c.f. Fig. \ref{fig:slopeslant}).  The contraction at large input delay is expected because the leading drop is the first to cross the bypass and slow down. Initially, it appears that expansion is not possible at large input delays. To further probe this notion, we developed an evolutionary algorithm \cite{Genetic_algo} to search for ladder designs containing any combination of slant and/or vertical bypasses that might be capable of contraction at low input spacing and expansion at large input spacing. Our search strategy revealed that such networks do exist, and an example is shown in Fig. \ref{fig:nonlinear}(a). The  dynamics of drop spacing in this network cannot be rationalized from mere addition of functionalities of the single bypasses shown in Fig. \ref{fig:slopeslant}. Instead, we find that the first five bypasses collectively cause contraction of drop spacing, while the last two bypasses cause drops to expand. However, the relative magnitudes of contraction and expansion from these sets of bypasses depends on the input drop spacing. We observe that at small input delays, the first five bypasses dominate, resulting in contraction behavior (see Fig. \ref{fig:nonlinear}(a)), whereas at large input delays, the last two bypasses dominate, yielding expansion.

A striking observation from Fig. \ref{fig:nonlinear}(a) is that the curve is significantly nonlinear compared to the almost linear dependence observed in ladders with identical bypasses. This result is significant because it implies nonlinear transformation of input delay without any droplet decision-making at bypass junctions. A unique consequence of this nonlinear transformation is the capability to encode and decode input delays as shown in Fig. \ref{fig:nonlinear}(b) where the entrance delay between pairs of drops represents the input signal and the system of bypasses represents the encoder and decoder. First consider the curves in Fig. \ref{fig:experimental} where the input signal is `scrambled' in the bypass section, but does not revert to its original value and therefore is not decoded. In striking contrast, we find in Fig. \ref{fig:nonlinear}(a), that input signal gets encoded and decoded at two different values of $\mathfrak{\Delta x_{in} = 4.5, 7(\Delta L)}$, where $\mathfrak{\Delta x_{in} = \Delta x_{out}}$. At these input delays, we find that the above-discussed contraction and expansion effects introduced by sets of bypasses negate each other. Thus, our results highlight a new route to code and decode signals compared to earlier studies that use drop decision making events in networks \cite{fuerstman}. The encoder/decoder illustration in this letter is just an exemplar demonstration of the potential of such networks in realizing complex behavior without any active components.

Given that ladders with mixed combinations of bypasses can display nonlinear behavior, we ask what degree of nonlinearity can be achieved in ladders. A qualitative indication can be obtained by considering reversibility in ladder networks. Reversibility implies that the original input delay is recovered when the flow is reversed. Ladder networks are reversible because of the absence of decision-making events \cite{ajdari}. This reversibility criterion demands that the relationship between input and output delays remain bijective. In addition, this functional relationship cannot have maxima, minima, or saddle points because it has to be strictly monotonic. Thus, we believe reversibility imposes bounds on the degree of nonlinearity that can be achieved with microfluidic ladders.

\indent In summary, we observe that MLNs with mixed bypasses display rich dynamics in drop spacing as well as nonlinear behavior. Such ladder networks in fact also exist in natural systems including leaf venation \cite{leaf_ven_reveiw}, microvasculature \cite{vascular_flow_review,vascular_flow_bmfc} and neural systems \cite{rat_ladder_neurons}. Interestingly, the nonlinear shape of Fig. \ref{fig:nonlinear}(a), resembles that of the widely observed sigmoid function, which is essential for coding/decoding neural signals \cite{single_neuron}, as it produces an invertible map. Finally, the framework described here can be expanded to explore not only pairs of drops, but also trains of drops to further probe collective hydrodynamics in drop-based microfluidic networks.\\
\indent We acknowledge National Science Foundation for partial financial support (Grant No. CDI-1124814).


\bibliography{apstemplate}
\bibliographystyle{apsrev4-1}

\end{document}